# Mapping of Secondary Virtual Networks onto Wireless Substrate based on Cognitive Radio: multi-objective formulation and analysis

Andson Marreiros Balieiro [1] and Kelvin Lopes Dias

Center of Informatics (CIn)
Federal University of Pernambuco (UFPE)
Email: {amb4, kld}@cin.ufpe.br

**Abstract:** There is a growing demand for wireless services with different requirements in a dense and heterogeneous wireless environment. Handling this complex ecosystem is becoming a challenging issue, and wireless virtualization emerges as an efficient solution. Although the inclusion of virtualization in wireless networks ensures a better use of resources, current approaches adopted for wireless virtualization can cause an underutilization of resources, since the resource allocated to a virtual wireless network is not shared with other one. This problem can be overcome by combining wireless virtualization with the cognitive radio (CR) technology and dynamic access spectrum (DSA) techniques. Thus, virtual wireless networks with different access priorities to resources (e.g., primary and secondary) can be deployed in an overlay form and share the same substrate wireless network, where the secondary networks use the resources opportunistically. However, challenges emerge in this new scenario, ranging from the mapping to the operation of these networks. This paper is the first to propose the cognitive radio virtual network environment and to formulate the problem of mapping the secondary virtual networks (SVNs) onto the wireless substrate network based on CR. To this end, a multi-objective formulation is designed. Moreover, an analysis of the metrics defined in the problem is performed in order to give the reader useful assistance in designing schemes to solve the problem of mapping SVNs onto substrate networks.

**Keywords:** Cognitive Radio Virtual Networks Environment; Secondary Virtual Networks Mapping; Primary Virtual Networks; Wireless Virtualization.

**1. Introduction**

Wireless standards and mobile device technologies are progressing quickly such that the future of digital communication will soon be dominated by a dense and heterogeneous wireless network [1]. Along with this, the demand for wireless services with different requirements in terms of security, quality of services (QoS) and quality of experience (QoE) is growing lately. In this respect, managing this complex wireless ecosystem is becoming a challenging issue, and wireless virtualization emerges as an efficient solution, since different virtual wireless networks can be created, sharing and running on the same wireless infrastructure, and providing differentiated services for users.

---
[1] Corresponding Address: Anibal Fernandes Avenue, 50740560, Recife, Pernambuco, Brazil. Tel.: +55 8198568611; E-mail address: amb4@cin.ufpe.br

Wireless virtualization involves both infrastructure and spectrum sharing, and introduces new players to the business model: the mobile network operator (MNO) and the service provider (SP). The former owns the network infrastructure and creates the virtual wireless networks, and the latter leases the virtual wireless networks, programs them and offers end-to-end services to users [2].

To satisfy the great demand for mobile communications, a natural and scarce resource – the electromagnetic spectrum – must be made available [3]. Although wireless virtualization provides a better use of resources, since the infrastructure is more widely shared, the current approaches adopted for wireless virtualization can cause an underutilization of resources, which is a problem.

In current approaches to virtualization, the resource allocated to a virtual network is not shared with another during the life time of the virtual networks. As their traffic load varies over time, there may be cases where the virtual wireless networks are not making full use of the resources allocated to them. This underutilization can have an adverse effect on the deployment of new virtual wireless networks and lead to a loss of revenue for the infrastructure provider, since the underutilized resources could be used for embedding new virtual networks.

Combining wireless virtualization with the cognitive radio (CR) technology and dynamic access spectrum (DSA) [3] techniques can overcome this problem of underutilization, where new virtual wireless networks can be deployed through opportunistic resource sharing and the deepest level of wireless virtualization (spectrum based virtualization [1]) can be achieved. This enables to have an environment composed of virtual wireless networks with different access priorities to the resources (e.g. primary and secondary), which are deployed in an overlay form and share the same substrate wireless network, in a situation where the virtual wireless network with lower priority (e.g. secondary) only has access to the resources when the higher priority network (e.g., primary) is not using them. With its cognitive and reconfiguration capabilities, the cognitive radio is able to sense the environment, detect the available spectrum bands (resources) and act on them, and thus enable the deployment of networks with different priorities. However, challenges emerge in this new scenario, ranging from the mapping to the operation of these networks.

Mapping virtual wireless networks onto wireless substrate networks is a NP-hard problem [4] and it involves reserving and allocating physical resources to elements that compose the virtual networks such as virtual base stations and virtual communication channels. This problem becomes more complex and challenging when it considers an environment composed of virtual wireless networks with different priorities of access to the resources, Primary Virtual Networks (PVNs) and Secondary Virtual Networks (SVNs), and that share common resources. The reason for this is that the SVN mapping must consider the demand requested by each SVN in terms of number of users (secondary users-SUs) and requested bandwidth, for example, in order to provide good quality of service to the SUs, and the usage pattern of the resources by the primary users, who are the users of the PVNs, to avoid causing harmful interference to PVN communication.

The mapping of SVNs imposes some constraints related to primary communication, such as keeping the interference in the PU communication below a defined threshold. It also seeks to focus on secondary communication, such as to minimize the handover of SU between SVNs and the blocking of SUs. Moreover, there are further aims with regard to the provider infrastructure, such as ensuring an efficient utilization of resources.

We can note that mapping SVNs is a multi-objective problem. Thus, a formulation, that takes into account these multiples objectives, needs to be performed. Moreover, it is important to analyze the metrics related to the objectives in order to know what can impair the performance of each objective and which objectives are conflicting with each other.

This paper formulates the problem of mapping the SVNs onto the substrate network as a multi-objective problem. It outlines the cognitive radio virtual network environment, which is formed by PVNs, SVNs and the substrate network. In this environment, the SVNs access the resources in an opportunistic way (only when the PVNs are not using them). Moreover, we discuss important metrics to be considered in the mapping process and perform an analysis of the influence of some parameters or metrics on other metrics in order to give the reader useful assistance in designing schemes to solve the problem of mapping SVNs onto substrate networks.

To the best of our knowledge, this is the first study to undertake the following: describe the cognitive radio virtual network environment, outline a formulation for the multi-objective problem of mapping virtual secondary networks onto substrate networks, considering the coexistence of the SVNs with the PVNs, and analyze the influence of some parameters and metrics on others.

This paper is organized as follows. Section 2 presents the related works. The cognitive radio virtual networks environment is described in Section 3. Metrics and the multi-objective problem of mapping SVNs onto substrate networks are formulated in Section 4. Section 5 conducts an analysis of the influence of some parameters/metrics on others. Section 6 concludes this paper.

**2. Related Works**

Several studies have been proposed for network virtualization in both wired [5] and wireless environments [6]. With regard to wireless networks, some studies have focused on virtualization considering a specific network technology [6][7], generic approaches, without specifying any adopted technology in the substrate network [8], or considering a scenario with heterogeneous wireless networks [9]. These studies believe that the resources allocated to a virtual network cannot be allocated to another network even if the first is not using it all the time. This restriction can cause resource underutilization, especially when the first virtual network experiences periods of low traffic load.

In light of this, in [5] the problem of embedding virtual networks in wired networks, based on an opportunistic sharing of resources, is raised. The authors consider the workload in a virtual network to be the combination of a basic sub-workload, which always exists, and a variable sub-workload, which occurs with some probability. Thus, multiple variable flows (traffic) from different virtual networks are allowed to share some common resources to achieve efficient resource utilization. However, this sharing can cause collisions/interference between variable traffic from different virtual networks.

In [10] the authors transfer the formulation given in [5] from the wired network scenario to wireless one. As in [5], in [10] the authors consider the resources to be homogeneous in terms of offered bandwidth. However, this does not encompass scenarios with heterogeneous networks (e.g. WiMax, LTE, WiFi), where different technologies have distinct resource units and different data rates [11]. Moreover, even with a single physical network technology, a factor such as noise can have an impact on the data rate achieved in a communication channel [12]. Unlike our proposal, in [5] and [10] the virtual networks keep the same level, regarding right of access to the resource, with no difference between primary virtual network/primary users (higher priority) and secondary virtual networks/secondary user (lower priority). In view of this, although the formulations in [5] and [10] are based on opportunistic resource sharing, they do not include these two elements of the cognitive radio environment: the secondary and primary networks/users, being [5] only focused on wired networks. In addition, in scenarios composed of virtual networks with different access priorities (e.g. primary and secondary), there are other factors (apart from the interference, represented by collision probability) that affect both user communication and resource utilization and that must be taken into account during the virtual networks mapping, such as the blocking probability of the SU and the handover probability of SUs between VNs, for example, which are neglected in both these studies.

A platform for end- to- end network virtualization is proposed in [13]. It aims to create virtual networks made up of wired and wireless resources, and uses CR to manage the virtualization in the wireless part. However, the authors only address how to abstract the heterogeneous wireless access networks for seamless connection with the wired virtual network. They neither formulate the problem of virtual networks mapping onto substrate networks, nor analyze the influences of the parameters/metrics on each other.

By using cognitive radio for wireless virtualization, in [14] the authors adopt an approach denoted as spectrum demand access as a service to achieve dynamic spectrum access (DSA). This approach dynamically offers spectrum services to users. They adopt the dynamic spectrum allocation approach for DSA, which does not distinguish between primary and secondary users. Thus, each user has a spectrum band for exclusive use within a certain time period. This type of exclusive allocation can cause spectrum underutilization when the traffic load is low. Moreover, the authors only consider homogeneous requests in their formulation, i.e. consider all virtual topologies request the same amount of spectrum, which is not always the case in real scenarios.

Unlike [14], our study adopts the opportunistic spectrum access (OSA) approach for DSA, where a difference is established between primary and secondary users. In OSA, the SUs dynamically search and access idle primary user spectrum bands through spectrum sensing or spectrum databases [15]. In view of this, in the secondary virtual network mapping, we take into account the existence of the primary virtual networks, which have common resources allocated to secondary networks and a higher access priority. In addition, we consider the heterogeneous demands (requests) in our formulation. In next section, we outline the environment considered in our multi-objective problem formulation.

## 3. The Cognitive Radio Virtual Networks Environment

The cognitive radio virtual network environment is made up of three types of wireless networks, substrate/infrastructure network, primary virtual networks (PVNs), and secondary virtual networks (SVNs), where each one is represented in Fig.1 and each network type is illustrated in a specific layer. The physical layer encompasses the substrate networks and consists of channels, spectrum bands, base stations, servers, and other features that compose the infrastructure of the wireless environment. The primary virtual networks are located on the primary virtual layer. The secondary virtual layer is formed by secondary virtual networks.

The substrate networks are used in the instantiation of both virtual networks (primary and secondary). In architectures like that proposed in [13], the infrastructure provider is the entity responsible for the management of the physical resources, which can cover different types of communication technology such as WiMax, WiFi, 3G/LTE. Moreover, these physical resources might belong to different infrastructure providers.

The PVNs have higher access priority to the resources (e.g. communication channels) than SVN. The PVN mapping is performed without taking account of the existence of the SVNs, in usual circumstances [8]. It does not take into account the concept of opportunistic sharing. Thus, the resources are divided and allocated to each PVN exclusively. As PVNs traffic load varies over time, there may be cases where the PVNs are not making full use of the resources allocated to them, which causes underutilization of the resources in the PVNs, especially in periods of low traffic load. This underutilization can lead to loss of revenue to the infrastructure provider, since the underutilized resources could be used for embedding new virtual networks.

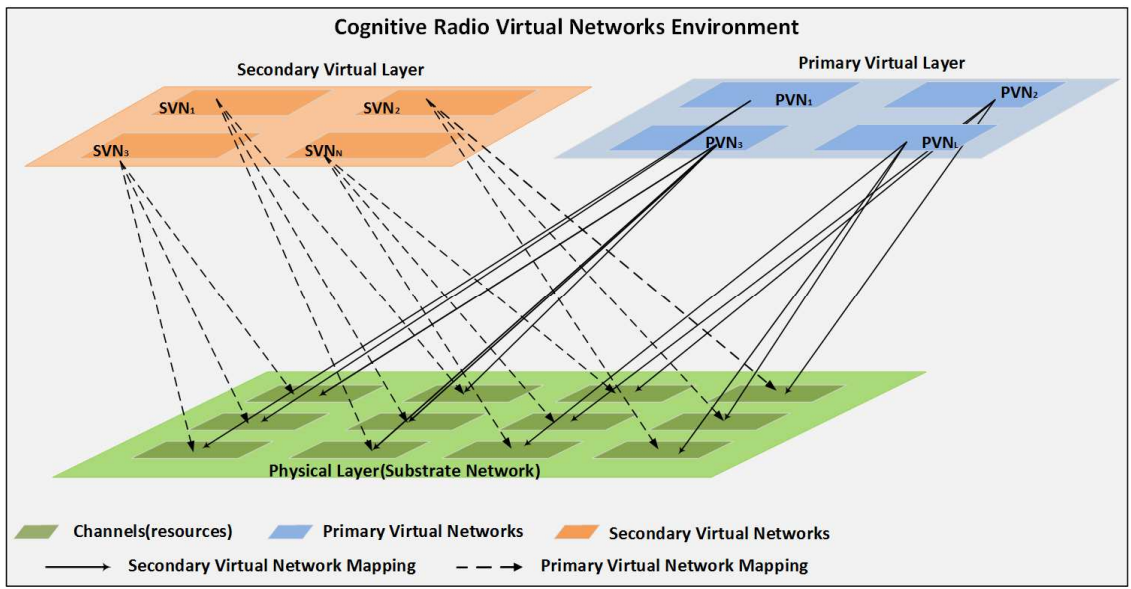

Figure 1. Cognitive Radio Virtual Networks Environment

Owing to the existence of low traffic periods in the PVNs, new virtual networks can be embedded through the opportunistic utilization of underutilized resources. These networks, denoted as secondary virtual networks, have lower access priority to the resources and only use them when the primary virtual networks are not using them. The adoption of the SVNs can provide better resource utilization (e.g. spectrum) and increased revenue for the provider infrastructure, because more virtual networks can be admitted.

However, in order to provide the opportunistic access to the resources, the elements of the SVNs must have mechanisms to identify the activity of the PVN users, denoted as primary user (PU) and access the resource during the absence of the PU, and release it when the PU returns. In this respect, the cognitive radio emerges as an essential technology for deploying the SVNs, due to its cognition and reconfiguration capabilities [3].

In an environment composed of virtual networks with different access priorities to resources (e.g. PVNs and SVNs), the virtual network mapping becomes more complex and challenging because the SVN mapping must take account of both the demand requested by the SVN, and the activity of the primary users, who are the users of the PVNs.

Thus, two points are important in the SVN mapping: awareness of higher access priority of the PVNs and their activities, to protect the PU from harmful interference caused by secondary users; to provide a good communication to SVNs.

## 4. The Problem of Secondary Virtual Networks Mapping Onto a Wireless Substrate Networks

This section presents the formulation of the problem of mapping secondary networks onto substrate networks, as a multi-objective problem. In this formulation, the primary and secondary virtual networks are seen as coexisting in the same substrate network. To the best of our knowledge, this is the first work that presents this formulation.

### 4.1 Formulation of the cognitive radio virtual network scenario

In a cognitive radio virtual networks scenario (adopting the two-level model [2]), the service provider requests the creation of and manages L primary virtual networks (PVNs). Given that the substrate network is composed of M channels (unit of resources), which are used for mapping the virtual networks, and that the mapping algorithm divides the resources between the PVNs according to percentage $q_j$, with $j=1,2,3,...,L$, $0 \leq q_j \leq 1$, and $\sum_{j=1}^{L} q_j = 1$, then for each PVN $j$ is allocated $|Q_j| = \lfloor M \cdot q_j \rfloor$ or $\lceil M \cdot q_j \rceil$ channels, where $Q_j$ means the set of channels allocated to PVN $j$, $\lceil x \rceil$ and $\lfloor x \rfloor$ are the ceiling and floor functions, respectively, aimed at keeping the following equality

$$\sum_{j=1}^{L} |Q_j| = M .$$

Given that the primary users arrival at channel $i(C_i)$ of the virtual network $j$, with $C_i \in Q_j$, follows a Poisson process with arrival rate $\lambda_{i,j}^{PU}$, and the user holding time given by an exponential distribution with mean $\frac{1}{\mu_{i,j}^{PU}}$, the utilization of channel $i$, according to [16], is given by Eq. 1.

$$\rho_{i,j}^{PU} = \left(\frac{\lambda_{i,j}^{PU}}{\mu_{i,j}^{PU}}\right) \quad (1)$$

An environment with opportunistic access of the SUs to the channels is provided by adopting the stability criterion $\rho_{i,j}^{PU} < 1$. Thus, the channels offer possibilities of opportunistic access by secondary users, and, hence, are amenable to be allocated to SVNs.

In this approach, in a similar way to [17], we abstract the existence of many primary users on the same channel. Thus, we consider each channel occupied by a PU per time at maximum. This includes the behavior of the many possible PUs on the channel. In this respect, the probability of the channel $i$ (that belongs to virtual network j) being busy (ON state) by PU is numerically equal to the utilization of the channel $i$, given by Eq.1.

On the other hand, the probability that channel $i$ is not being used by the PU, (i.e. it is in the OFF state) can be denoted as Eq.2.

$$POFF_{i,j} = (1 - \rho_{i,j}^{PU}) \quad (2)$$

Thus, the average number of idle channels (OFF state) in the virtual network $j$ can be given by Eq.3.

$$E[NCh_j^{OFF}] = \sum_{n=1}^{|Q_j|} n \cdot P[NCh_j^{OFF} = n] \quad (3)$$

Where $P[NCh_j^{OFF} = n]$ denotes the probability that there are $n$ channels in the OFF state (no PUs) in the virtual network $j$, with $n = 0,1,2,3,...,|Q_j|$.

Let $A_{d,n,j}$ be the partition $d$ with $n$ elements of the set $Q_j$, with $d = 1,2,3...,s$, where $s$ is the number of partitions with $n$ elements of the set $Q_j$, given by $s = \binom{|Q_j|}{n}$, the value of $P[NCh_j^{OFF} = n]$ can be obtained by Eq.4.

$$P[NCh_j^{OFF} = n] = \sum_{d=1}^{s} \left[ \prod_{i \in A_{d,n,j}} POFF_{i,j} \cdot \prod_{h \in \{Q_j - A_{d,n,j}\}} [1 - POFF_{h,j}] \right] \quad (4)$$

In a similar way, the probability that there are $n$ primary users in the virtual network that was mapped on the set of channels $Q_j$ can be obtained from Eq. 5.

$$P[NPU_j = n] = \sum_{d=1}^{s} \left[ \prod_{i \in A_{d,n,j}} \rho_{i,j}^{PU} \cdot \prod_{h \in \{Q_j - A_{d,n,j}\}} [POFF_{h,j}] \right] \quad (5)$$

The wireless communication is influenced by several factors such as noise from other sources, weather, and obstacles in the environment. These factors affect the signal strength perceived by receptor, which varies during the transmission. Thus, to denote the variations in the primary channel conditions, we consider that the signal- to-noise ratio perceived in channel $i$ of the virtual network $j$, $SNR(db)_{i,j}$, is given by an exponential distribution with mean $\frac{1}{\lambda_{i,j}^{SNR(db)}}$ dB [18]. Thus the maximum capacity (in bps) of the channel $i$ allocated to the virtual network $j$, denoted as $R_{i,j}$, is given by Shannon´s law [19] and expressed in Eq.6, where $Bw_{i,j}$ means the channel bandwidth (in Hertz).

$$R_{i,j} = Bw_{i,j} \cdot \log_2(1 + SNR_{i,j}) \quad (6)$$

The value of $SNR_{i,j}$ using the decimal scale obtained from the SNR in decibels (dB) is given by Eq.7.

$$SNR_{i,j} = 10^{\frac{SNR(db)_{i,j}}{10}} \quad (7)$$

The average capacity achieved on channel $i$ of the virtual network $j$, denoted as $\overline{R_{i,j}}$, can be calculated by means of Eq. 8.

$$\overline{R_{i,j}} = E[R_{i,j}] = E[Bw_{i,j} \cdot \log_2(1 + (SNR_{i,j}))] \quad (8)$$

Eq.8 represents the average channel capacity when it is always available for use by the user, as occurs when the user is the PU. When the user is the SU, this does not occur. The SU only uses the channel in an opportunistic way, when the PU is absent. Thus, considering Eq.8 and the OFF state probability, given by Eq.2, the effective rate (on average) achievable through the opportunistic use on channel $i$ of the virtual network $j$ is given by Eq. 9.

$$\overline{Re_{i,j}} = POFF_{i,j} \cdot \overline{R_{i,j}} \quad (9)$$

The SVNs provide their services by using the resources allocated to the PVNs in an opportunistic way. In virtual network mapping, there is no one-to-one relationship between PVNs and SVNs in this environment, i.e. each SVN only has to be mapped over one PVN and vice-versa. Thus, channels allocated to different PVNs can be used by the same SVN, as shown in Fig. 1, where SVN #3 uses the channels allocated to PVN #1 and PVN #3, for example, which provides broader network mapping and more possibilities for SVN mapping. This is unlike [20] and [21] which adopt an approach that is restricted to a one-to-one relationship between SVNs and PVNs. Thus, in the following formulations, we adopted 'the channel' as level of granularity, when information about primary virtual activity is taken into account.

For the secondary virtual layer, in this approach, we consider that there are $N$ SVNs to be mapped onto the substrate network. In each $SVN\ l$ ($SVN_l$), with $l=1,2,3..,N$, the SUs perform applications with data rate/bandwidth (BW) requirements given by an exponential distribution with mean $1/w_l$ bps. The arrival of SUs on the $SVN\ l$ follows a Poisson distribution with a rate of $\lambda_l^{SU}$ users per second. The SU holding time is given by an exponential distribution with mean $1/\mu_l^{SU}$ second [22]. Thus, the average number of SUs in the $SVN\ l$, and the amount of resources requested by SUs can be calculated by Eq. 10 and Eq.11, respectively.

$$\overline{NSU_l} = \lambda_l^{SU} \cdot \frac{1}{\mu_l^{SU}} \qquad (10)$$

$$\overline{Bw_{req,l}} = \lambda_l^{SU} \cdot \frac{1}{\mu_l^{SU}} \cdot \frac{1}{w_l} = \overline{NSU_l} \cdot \frac{1}{w_l} \qquad (11)$$

**4.2 Formulation for Collision Probability**

In the SVNs mapping, it is important to consider other factors apart from the demand for these networks. As the channels adopted in this mapping are shared with the PVNs, which have higher access priority, it is necessary to ensure that the interference level caused to primary communication does not go beyond a defined threshold. This threshold can be defined on the basis of the service level agreement (SLA)/service level specification (SLS) from the PVNs or the interference level that can be tolerated by the applications/signals of the PVNs, for example. Thus, in selecting the channels that must be allocated to each SVN, the interference or collision probability between PU and SU must be calculated to ensure that its value will be below the defined threshold.

Given that the mapping of the SVN $l$ onto the substrate network adopted the set of channels, $SC_l = \{C_1, C_2, ..., C_n\}$, where $|SC_l| = n_l$, $SC_l \subset \bigcup_{j=1}^{L} Q_j$, and $SC_l \cap SC_u = \varnothing$, for all $l \neq u$, with $l, u = 1, 2, 3, ..., N$ being identifiers of SVNs. In the virtual networks mapping (secondary or primary), when each user (SU or PU) needs only one channel to perform its communication, i.e. to meet its demands in terms of data rate, the collision between PU and SU on the $SVN\ l$ can occur when the number of PUs and SUs that attempt to access the channels simultaneously are greater than the number of channels allocated to $SVN\ l$

(13)

, with at least one PU and one SU between the users. Thus, the collision probability between PU and SU on the *SVN l* is given by Eq.13.

$$Pc_l(SC_l) = \sum_{i=1}^{n_l} (P[NPU_l = i \ ; \ NSU_l > n_l - i])$$

Where $P[NPU_l = i \ ; \ NSU_l > n_l - i]$ is the probability that in the set of channels $SC_l$ allocated to *SVN l*, the number of PUs in the channels or seeking to access the channels is $i$ and the number of SUs is greater than $n_l - i$. As the network mapping/dimensioning takes place before the operation of the network, the users are not using the system yet. Their arrival and access attempts occur independently of each other. Thus, $P[NPU_l = i \ ; \ NSU_l > n_l - i]$ can be obtained from the product between two probabilities, $P[NPU_l = i]$ and $P[NSU_l > n_l - i]$. The first probability is given by Eq.5, and the second by Eq.14.

$$P[NSU_l > n_l - i] = 1 - \left( \sum_{k=0}^{n_l - 1} P[NSU_l = k] \right) \tag{14}$$

Where

$$P[NSU_l = k] = \frac{\left( \lambda_l^{SU} \cdot \frac{1}{\mu_l^{SU}} \right)^k e^{-\left( \lambda_l^{SU} \cdot \frac{1}{\mu_l^{SU}} \right)}}{k!} \tag{15}$$

However, in this formulation, the condition that each user needs only one channel to meet its demand, does not apply to the SU, but just to the PU, which we considered that the PU acts on one channel. As the bandwidth requested by the SU may be greater than the average capacity of a channel, this user will need more than one channel to perform its communication. When the SU tries to access more than one channel on the SVN, it may collide with more than one PU. The average demand requested by each SU and the average capacity offered by the channels from the set $CM_l$ must be taken into account before the collision probability on the *SVN l* can be estimated. Thus, the collision probability can be obtained by Eq.16.

$$Pc_l(SC_l) = \sum_{i=1}^{n} \left( P\left[ NPU_l = i \ ; \ NSU_l > \left\lfloor \frac{n_l - i}{ChSU_l} \right\rfloor \right] \right) \tag{16}$$

Where $ChSU_l$ is the average number of channels required to meet the demand of each SU on the *SVN l*, which is given by Eq.17, and $\lfloor . \rfloor$ is the floor function.

$$ChSU_l = \max\left( 1, \frac{1/w_l}{Rch_l} \right) \tag{17}$$

With $\overline{Rch_l}$ being the average effective capacity of the channels allocated to $SVN\ l$, given by Eq.18, and $1/w_l$ the average data rate requested by each SU on this network.

$$\overline{Rch_l} = \frac{\sum_{i=1}^{n_l} \overline{Re_{i,l}}}{n_l} \qquad (18)$$

When the collision probability of each SVN is obtained, the average collision probability on the secondary virtual layer, if all the N mapped SVNs are considered, is given by Eq. 19.

$$\overline{Pc} = \frac{\sum_{l=1}^{N} Pc_l}{N} \qquad (19)$$

**4.3 Formulation for Blocking Probability**

As well as giving protection to the primary users by restricting the collision probability, the mapping of SVNs onto the substrate network must provide good quality of service for the SUs. Thus, it must admit as many SUs as possible dimensioned for each SVN. For this reason, the SU blocking probability needs to be determined in the mapping process. The SU blocking occurs in the $SVN\ l$ when the total number of PUs together with the average number of channels requested by SUs, is greater than the number of channels allocated to $SVN\ l$. Hence, the SU blocking probability on the $SVN\ l$ is calculated by Eq.20.

$$Pb_l^{SU}(SC_l) = \sum_{i=0}^{n} (P[NPU_l = i\ ;\ NSU_l > \left\lfloor \frac{n_l - i}{ChSU_l} \right\rfloor]) \qquad (20)$$

When the mapping of N SVNs is considered, the SU blocking probability on the secondary virtual layer (on average) is given by Eq.21.

$$\overline{Pb^{SU}} = \frac{\sum_{l=1}^{N} Pb_l^{SU}}{N} \qquad (21)$$

**4.4 Formulation for Joint Utilization**

As well as seeking to admit as many users as possible, providing better resource utilization (e.g. better utilization of channels) is also a goal of both SVN mapping and cognitive radio technology through opportunistic access to the resources. Before defining the resource utilization achieved by a given SVN mapping that uses channels shared with PVNs, the activities of both the SVN and PVNs must be considered. Thus, given the set of $n_l$ channels $SC_l = \{C_1, C_2, ..., C_{nl}\}$ used in the mapping of the $SVN\ l$, the joint utilization (primary and secondary usage) of these channels is given by Eq.22, which is the ratio of the average number of channels occupied by PUs or SUs to the number of channels allocated to *SVN l* .

$$util_l = \frac{\overline{NPU}_l + (1 - Pb_l^{SU}).\overline{NSU}_l.ChSU_l}{n} \qquad (22)$$

Where $\overline{NPU}_l$ is the average number of PU in channels shared with *SVN l*, calculated by Eq. 23, which is the difference between the number of channels allocated to *SVN l* and the average number of channels in the OFF state, without PUs, expressed in Eq.3.

$$\overline{NPU}_l = n_l - E[NCh_l^{OFF}] \qquad (23)$$

The product $(1 - Pb_l^{SU}) * \overline{NSU}_l$ denotes the average number of SUs admitted in the *SVN l*, where $\overline{NSU}_l$ is given by Eq.10, and $Pb_l^{SU}$ by Eq. 20.

In the SVN mapping, it is important to avoid resource underutilization, such as spectrum, which is a scarce natural resource, and to increase the revenue of the infrastructure provider. When average resource utilization is provided by each SVN, the average joint utilization achieved by mapping of *N* SVNs in conjunction with the utilization provided by primary virtual networks is given by Eq.24.

$$\overline{util} = \frac{\sum_{l=1}^{N} util_l}{N} \qquad (24)$$

**4.5 Formulation for SVN Handover Probability**

The SU blocking probability given in Eq.20 means the rejection level of new SUs in the SVN, i.e. the network capacity of admit/reject new secondary users.

Once the SUs are admitted, some events triggered by PU activity can affect the quality of service offered to them. Among these events, the spectrum (channel) handover and secondary virtual networks handover can be highlighted. They occur as a result of the return of the PU to the channel occupied by SU. In the spectrum handover, the channel selected by SU to resume its communication is available in its current SVN. On the other hand, in the SVN handover, the SU selects a channel in another SVN, since there is no available channel in its current SVN. Thus, the SU has to change the channel and the SVN before it can resume its communication.

The SVN handover causes greater degradation to secondary communication than spectrum handover, because it results in signaling overhead and increases the processing delay of ongoing connections with handover [20].

As presented, the SVN handover occurs when the SU switches between channels from different SVNs, due to the appearance of PU in its communication channel, and when there are not enough available channels in its current SVN. Thus, before the SVN handover can be triggered, two conditions have to be satisfied: the preemption of the SU from its current network and the existence of enough available channels in any other SVN. It is important to note that the SU preemption from *SVN l* occurs when the SU collides with the PU in its current SVN and it there are not enough available channels to allow the SU to resume its communication. The handover probability of the SU from the *SVN l* can be obtained by the ratio between the number of successful handover attempts (those that have found enough available resources in other SVNs) and the number of SUs admitted in the *SVN l*.

Let $Pb_l^{SU}$ be the SU blocking probability in the *SVN l*, the number of SU admitted in this virtual network ($NAd_l^{SU}$) is given by Eq. 25.

$$NAd_l^{SU} = (1 - Pb_l^{SU})\lambda_l^{SU} \cdot \frac{1}{\mu_l^{SU}} \tag{25}$$

From the admitted users in the *SVN l*, some can try to execute SVN handover, and these are the ones that collide with the PU. Thus, the number of SVN handover attempts, denoted as $NHA_l$, is given by Eq.26.

$$NHA_l(SC_l) = PcAd_l \cdot NAd_l^{SU} = PcAd_l \cdot (1 - Pb_l^{SU})\lambda_l^{SU} \cdot \frac{1}{\mu_l^{SU}} \tag{26}$$

Where $PcAd_l$ is the collision probability between the SUs admitted in the *SVN l* and the PUs, which can be calculated in a similar way to Eq. 16, where the probability $P[NSU_l = k]$, given in Eq.15, is changed to the following, given in Eq. 27.

$$P[NSU_l = k] = \frac{\left((1 - Pb_l^{SU})\lambda_l^{SU} \cdot \frac{1}{\mu_l^{SU}}\right)^k e^{-\left((1-Pb_l^{SU})\lambda_l^{SU} \cdot \frac{1}{\mu_l^{SU}}\right)}}{k!} \tag{27}$$

The SVN handover attempt probability of the SUs accepted in the *SVN l* is denoted as Eq.28, which is numerically equal to the collision probability between the SUs accepted in the *SVN l* and the PUs

$$P_{HA,l}(SC_l) = \frac{NHA_l}{NAd_l^{SU}} = PcAd_l \tag{28}$$

There may not be enough available resources in the other SVNs to admit all the SUs that performed handover from the *SVN l*. Thus, the number of successful SVN handovers depends on the amount of available resources in the other SVNs.

Since there are (on average) R available resources in the other SVNs, the number of successful SVN handovers is given by Eq.29.

$$NH_l = \frac{Min\ (NHA_l \cdot ChSU_l^*, R)}{ChSU_l^*} \tag{29}$$

Where $ChSU_l^*$ is the average number of necessary channels to continue the communication of each SU that tries to perform handover from $SVN\ l$ to another SVN. This number can be obtained in a similar way as Eq. 17, by considering the average effective capacity of the channels allocated to the other SVNs, with the exception of the $SVN\ l$, as denominator. The R value is given by Eq. 30.

$$R = \max\left(0, \sum_{j=1, j \neq l}^{N} \left[ n_j - \left( \overline{NPU}_j + (1 - Pb_j^{SU}) \cdot \overline{NSU}_j \cdot ChSU_j \right) \right] \right) \tag{30}$$

Thus, considering the SUs admitted in the $SVN\ l$, the SVN handover is obtained by Eq. 31.

$$P_{handover,l} = \frac{NH_l}{NAd_l^{SU}} \tag{31}$$

Taking into account the mapping of N SVNs, the average SVN handover probability, in the secondary virtual layer, is denoted by Eq. 32.

$$\overline{P_{handover}} = \frac{\sum_{l=1}^{N} P_{handover,l}}{N} \tag{32}$$

**4.6 The Multi-objective Problem of SVNs Mapping onto Substrate Networks**

As shown above, several objectives must be considered in the SVN mapping process. When the influence of each of these on the SU and PU communications is taken into account, the following optimization problem can be formulated (see Eq.33). Given a set of $N$ SVN requests and the usage pattern of the channels (resources) for the primary networks, to perform the mapping of the SVNs in order to minimize the SVN handover and the SU blocking probabilities and maximize the joint utilization. Three constraints must be overcome: the collision probability in each SVN must be below a desired threshold, the amount of resources allocated to each SVN cannot be less than the requested demand, and a common channel cannot be allocated to different SVNs. In notation, we have:

$$\begin{aligned} & Minimize\ \overline{P_{handover}},\ \overline{Pb^{SU}}\ and\ Maximize\ \overline{util} \\ & Subject\ to: \\ & \quad \begin{cases} Pc_l < thr_{collision},\ l = 1,2,3,...,N \\ Ralloc_l \geq \overline{Bw_{req,l}},\ l = 1,2,3,...,N \\ SC_l \cap SC_u = \emptyset,\ l \neq u,\ l,u = 1,2,3,...,N \end{cases} \end{aligned} \tag{33}$$

Where $Ralloc_l$ and $SC_l$ are the amount of resources and the set of allocated channels to $SVN\ l$, respectively, and $thr_{collision}$ is the collision probability threshold.

**5. Analysis of the Secondary Virtual Networks Mapping**

This section conducts an analysis of some of the metrics and parameters defined in the previous section. It aims to show the influence of some metrics and parameters on other metrics and give the reader useful assistance in designing schemes to solve the problem of mapping SVNs onto substrate networks as defined in previous section.

There are some constraints in the SVN mapping problem, which were outlined in the previous section, like the demands requested by SVN in terms of throughput, collision probability threshold, and objectives (such as achieving low blocking and SVN handover probabilities, and high resource utilization). Some of these constraints and objectives concern primary communication, such as keeping the collision probability (interference to PU) below a defined threshold, while others involve secondary communication, such as the question of how to minimize the SVN handover and SU blocking probabilities. In addition, there is, for example, the question of how the provider infrastructure and network environment can provide efficient resource utilization.

Achieving all these objectives simultaneously is a challenging process, because some conflict with each other. If the SVN mapping is focused on a specific objective, it can deteriorate other one.

Thus, in this section an analysis is conducted of some metrics to achieve certain objectives, when a particular parameter varies. This analysis aims to provide an overview of the impact of certain parameters on some metrics (objectives), and of one objective on another objective.

The primary utilization rate of the channels is an important parameter and must be taken into account in the SVNs mapping, because it directly affects metrics like the collision and SU blocking probabilities and joint utilization of resources, which are among the objectives of the problem defined in Eq. 33.

Fig 2 considers the number of channels allocated to SVN, SU arrival rate and SU holding time to be fixed values, and shows the behavior of the collision probability, SU blocking probability and joint resource utilization when the primary utilization rate of the channels allocated to a particular SVN varies.

Low SU blocking and collision probabilities can be achieved by selecting the channels with a low primary utilization rate to map the SVN, as illustrated in Fig. 7. However, choosing channels with a low primary utilization rate can cause a reduction in the joint utilization of the resources, despite the higher SU admission rate in the SVN, which is inferred by a lower SU blocking probability and higher resource utilization by SU. On the other hand, if channels with a high primary utilization rate are selected, higher resource utilization can be achieved, as illustrated in Fig.7. But, at the same time, this might cause an increase in the collision and SU blocking probability, which will affect both the primary and secondary communications.

In addition, although there is an increase in the joint utilization when channels with high primary utilization are selected, the resource utilization by secondary users declines until it reaches the null value, and there is no longer any possibility of the SUs using the selected channels opportunistically, i.e., only the PUs use the channels, as illustrated in Fig.2.

Thus, in the mapping of a SVN, it is important to balance the selection of the channels, by not just adopting channels with a high primary utilization rate or a low rate, to achieve a good tradeoff between the collision probability, SU blocking probability and joint resource utilization.

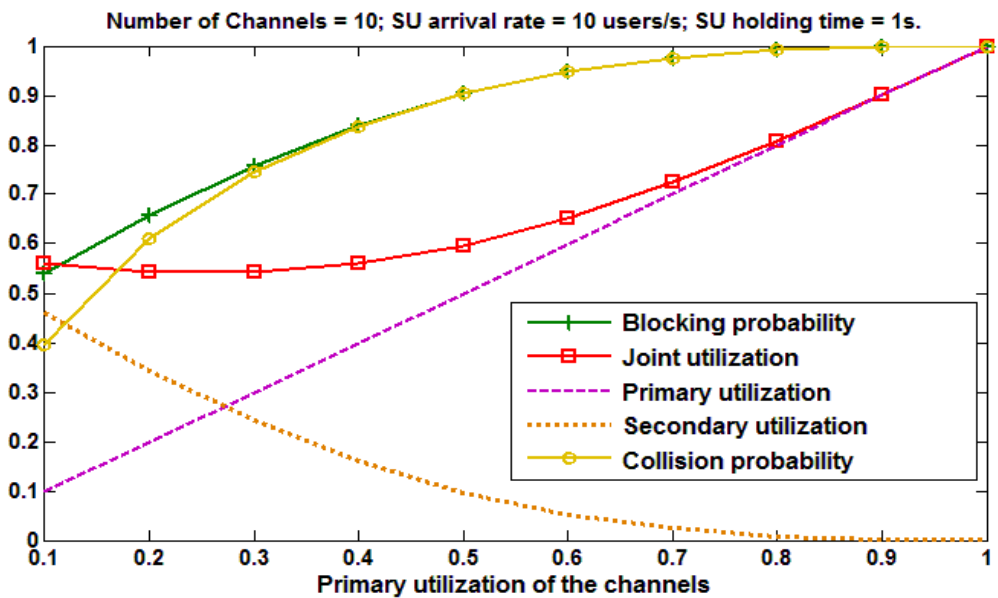

Figure 2. Influence of Primary Utilization on SVN mapping

Fig 3 takes into account the number of channels adopted in the mapping of the SVN, and illustrates the results for the collision, SU blocking and handover attempt probabilities, and joint utilization, when the number of allocated channels varies and the primary utilization rate, SU arrival rate and SU holding time are fixed. It can be noted that the collision and SU blocking probabilities decline when the number of channels allocated to SVN increases. This means that there is less interference to PU communication and fewer SUs are rejected in the SVN, which are positive factors. In addition, with more channels allocated to SVN, there are generally more resources to meet the SU demand and, hence, less chance of the SU performing handover. However, the increase in the number of channels, from a given point (14 channels), usually leads to a reduction in joint resource utilization, as illustrated in Fig.3, because despite the low blocking rate of SU in the SVN, the denominator in Eq.22 grows faster than the numerator, i.e. the amount of allocated resources to SVN is much larger than the demands to be met.

On the other hand, if a reduced number of channels is employed in the SVN mapping, the joint resource utilization increases, which is a positive factor, but the collision and SU probabilities increase too, which is a negative one.

It is worth noting the behavior of the SVN handover attempt/ probability, which composes the SVN handover probability, and joint resource utilization, when the number of channels varies within a particular range, (between 10 and 14). Within this range, when there is an increase in the number of allocated channels, the joint resource utilization and the handover attempt probability increase too. Given the joint resource utilization, when there were more channels, more SUs were admitted and the numerator of Eq.20 grew more than the denominator. With regard to the handover attempt probability, the increase of the number of channels caused a higher admission of SU in the SVN, but the number of channels did not grow proportionality to the admission of SUs. Thus, the SVN occupation rate increased, which increased the likelihood of the SU switching from the SVN.

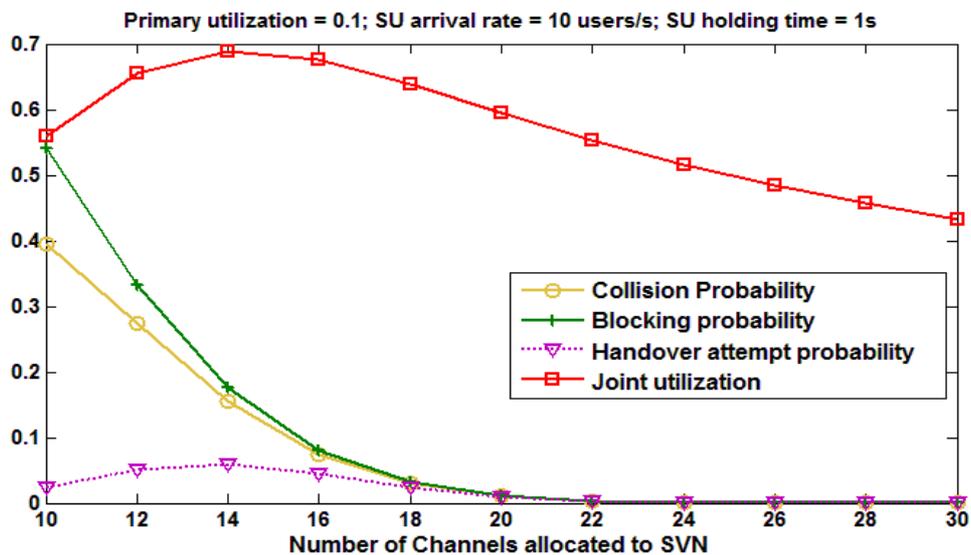

Figure 3. Analysis of variations in the number of Channels Allocated to SVN

On the basis of the previous analysis, it should be pointed out that the blocking probability has an impact on both joint resource utilization and handover attempt probability. The blocking probability indicates the level of admission/rejection of SU in the SVN. Thus, Fig.4 illustrates the behavior of the handover attempt probability and joint resource utilization when different values are defined for the blocking probability. When the blocking probability increases, the joint resource utilization and the handover attempt probability decrease, because fewer SUs are admitted in the SVN. Thus, in selecting channels that will be allocated to SVN, the influence of the blocking probability on other metrics should be taken into account.

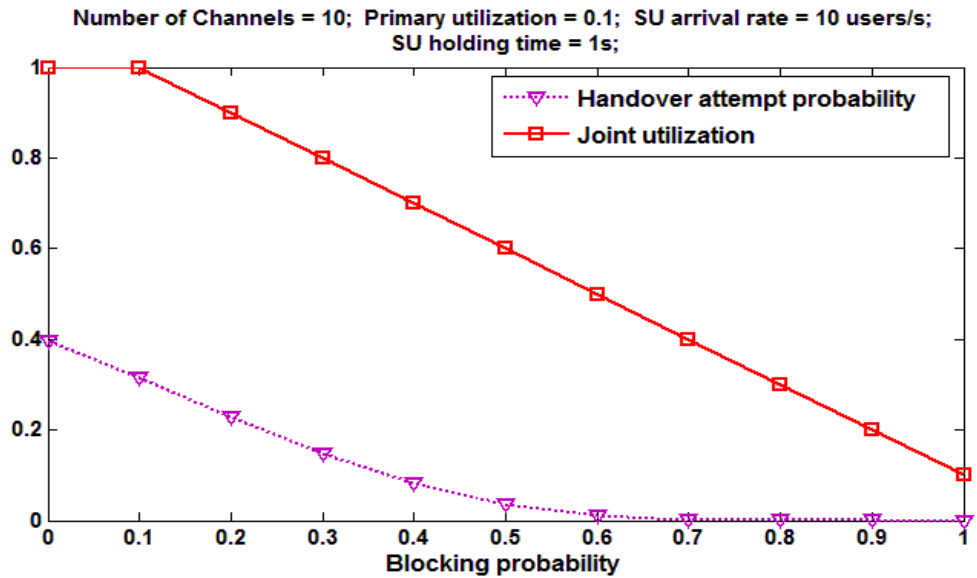

Figure 4. Influence of the blocking probability

**6. Conclusions and Future Directions**

In this paper, we have addressed the combination of cognitive radio, dynamic spectrum access techniques and wireless virtualization in order to overcome the problem of resource underutilization that can occur when current approaches to wireless virtualization are adopted. Although this combination enables virtual wireless networks with different priorities of access to the resources (PVNs and SVNs) can share common substrate networks, where the SVNs access the resources opportunistically, the mapping of SVNs onto substrate networks is a challenging problem, where there are objectives related to the PU, SU and MNO. Thus, we have outlined this new scenario and formulated the SVNs mapping onto substrate networks as a multi-objective problem. Moreover, an analysis of the influences of some parameters/metrics on other metrics was performed. It aimed to give the reader useful assistance in designing schemes to solve the problem of mapping SVNs onto substrate networks.

Future works includes designing a scheme to solve the SVN mapping problem and evaluating it in terms of metrics as collision probability, SU blocking probability, SVN handover probability and joint resource utilization.

## 7. Acknowledgments

This work was partially supported by a grant from the Science and Technology Foundation of Pernambuco (FACEPE)/Brazil.Reg. Nº IBPG -0571-1.03/10.

## 8. References

[1] H. Wen, P. K. Tiwary, T. Le-Ngoc, "Current trends and perspectives in wireless virtualization", International Conference on Selected Topics in Mobile and Wireless Networking (MoWNeT), 2013.

[2] C. Liang, F. R. Yu, "Wireless Network Virtualization: A Survey, Some Research Issues and Challenges", IEEE Communications Survey and Tutorials, 2014.

[3] I. F. Akyildiz, W. –Y. Lee, M. C. Vuran and S. Mohanty, "Next generation/dynamic spectrum access/cognitive radio wireless networks: A survey", Elsevier Computer Networks (50), 2006.

[4] A. Belbekkouche, M. M. Hasan, A. Karmouch, "Resource Discovery and Allocation in Network Virtualization", IEEE Communications Surveys & Tutorials, Vol.14, Issue 4, pp.1114-1128, 2012.

[5] S. Zhang, Z. Qian, J. Wu, S. Lu, "An Opportunistic Resource Sharing and Topology-Aware mapping framework for virtual networks", Proceedings of IEEE INFOCOM, 2012.

[6] Y. Zaki, L. Zhao, C. Goerg, A. Timm-Giel, "LTE wireless virtualization and spectrum management", Third Joint IFIP Wireless and Mobile Networking Conference, 2010.

[7] A. Banchs, P. Serrano, P. Patras, M. Natkaniec, "Providing Throughput and Fairness Guarantees in Virtualized WLANs Through Control Theory", Springer Mobile Networks and Applications, vol.17, issue 4, pp 435-446, 2012.

[8] F. Fu, U. C. Kozat, "Wireless Network Virtualization as A Sequential Auction Game",Proceedings of IEEE INFOCOM, 2010.

[9] L. Caeiro, F. D. Cardoso, L. M. Correia, "Adaptive allocation of Virtual Radio Resources over heterogeneous wireless networks", 18th European Wireless Conference (European Wireless), 2012.

[10] M. Yang, Y. Li, D. Jin, J. Yuan, L. Su, L. Zeng, "Opportunistic Spectrum Sharing Based Resource Allocation for Wireless Virtualization", Seventh International Conference on Innovative Mobile and Internet Services in Ubiquitous Computing (IMIS), 2013.

[11] X. Wang, P. Krishnamurthy, D. Tipper, "Wireless Network Virtualization", International Conference on Computing, Networking and Communications, 2013

[12] Yuh-Shyan Chena, Ching-Hsiung Choa, I. Youb, Han-Chieh Chao, "A Cross-Layer Protocol of Spectrum Mobility and Handover in Cognitive LTE Networks", Elsevier Simulation Modelling Practice and Theory, vol. 19, issue 8, pp.1723–1744, 2011.

[13] K. Nakauchi, K. Ishizu, H. Murakami, A. Nakao, H. Harada "AMPHIBIA: A Cognitive Virtualization Platform for End-to-End Slicing", IEEE International Conference on Communications (ICC), 2011.

[14] C. Xin, M. Song, "Dynamic Spectrum Access as a Service", Proceedings IEEE INFOCOM, 2012.

[15] A.W.Min,K-H. Kim,J. P.Singh,K. G. Shin, "Opportunistic spectrum access for mobile cognitive radios", Proceedings of IEEE INFOCOM, pp. 2993-3001, 2011.

[16] G. Bolch, S. Greiner, H. De Meer, K. Trived, "Queueing Networks and Markov Chains", Second Edition, Wesley, 2006.

[17] L. Akter, B. Natarajan, C. Scoglio, "Modeling and Forecasting Secondary User Activity in Cognitive Radio Networks", Proceedings of 17th International Conference on Computer Communications and Networks (ICCCN),pp. 1-6, 2008.

[18] S. Choudhury, J. D. Gibson, "Information Transmission Over Fading Channels", IEEE Global Telecommunications Conference (GLOBECOM), pp. 3316-3321, 2007.

[19] C.E. Shannon, "A Mathematical Theory of Communication", Association for Computing Machinery, Vol. 5, No. 1, pp. 3-55, March 2001.

[20] J. Lai, E. Dutkiewicz, R. P. Liu, R. Vesilo, G. Fang, "Network Selection in Cooperative Cognitive Radio Networks", 11th International Symposium on Communications and Information Technologies (ISCIT),2011.

[21] C. Wang, K. Sohraby, R. Jana, L. Ji, M. Daneshmand, "Network Selection in Cognitive Radio Systems", IEEE Global Telecommunications Conference, (GLOBECOM), 2009.

[22] L. Zhao, M. Li, Y. Zaki, A. Timm-Giel, C. Gorg, "LTE Virtualization: from Theoretical Gain to Practical Solution", 23rd International Teletraffic Congress (ITC), 2011.